\documentclass[aps,prl,multicol,psfig]{revtex}

\usepackage{graphics}
\input epsf

\begin{document}

\title{On the Friedmann Equation in Brane-World Scenarios}

\author{Justin Khoury$^1$ and Ren-Jie Zhang $^2$}

\address{$^1$Joseph Henry Laboratories,
Princeton University,
Princeton, NJ 08544, USA \\
$^2$ School of Natural Sciences, Institute for Advanced Study,
Princeton, NJ 08540 USA}

\maketitle

\begin{abstract}

The Friedmann law on the brane generically depends quadratically on the brane energy density and involves a ``dark radiation'' term due to the bulk Weyl tensor.
Despite its unfamiliar form, we show how it can be derived from a standard four-dimensional Brans-Dicke theory at low energy.
In particular, the dark radiation term is found to depend linearly on the brane energy densities.
For any equation of state on the branes, the radion evolves such as to generate radiation-dominated cosmology.
The radiation-dominated era is conventional and consistent with nucleosynthesis.

\end{abstract}

\begin{multicols}{2}[]

There has been considerable interest recently for models where space-time is effectively five-dimensional, with the extra dimension compactified on an $S_1/{\bf Z_2}$ orbifold.
Coincident with the fixed planes of the ${\bf Z_2}$ symmetry are two 3-branes (or boundary branes).
Gauge and matter fields are confined to the boundary branes, while gravity and other bulk fields can propagate in the whole of space-time.
This general set-up is motivated by heterotic M-theory~\cite{witten,burt} and the Randall-Sundrum (RS) model~\cite{randall}.

For simplicity, in the present work we assume that the bulk is empty.
It is known that the induced Friedmann law on either brane in this case is given by~\cite{shiro,bine2,shinji}
\begin{equation}
3H_i^2 = \frac{{\cal C}}{a_i^4} + \frac{\rho_i^2}{12}\,,
\label{eq:main}
\end{equation}
where $a_i$ and $H_i$ are respectively the induced scale factor and Hubble parameter on the brane in question, while $\rho_i$ denotes its energy density. 
(In this paper, the five-dimensional Planck mass is unity.)
The ${\cal C}/a_i^4$ term, where ${\cal C}$ is an arbitrary constant, originates from the projection of the bulk Weyl tensor onto the brane.

The above equation (along with energy conservation for $\rho_i$) is sufficient to determine the time evolution of $a_i$.
However, to fully specify the cosmology, we must also understand the nature of the energy content driving the expansion. 
For instance, if ${\cal C}/a_i^4$ described a new form of dark energy, then Eq.~(\ref{eq:main}) would imply that $H_i^2$ depends quadratically on the matter density, rather than linearly as in four-dimensional cosmology~\cite{lukas2,bine1}.
This would be inconsistent with nucleosynthesis.
Moreover, it would contradict the well-established principle that physics should appear four-dimensional at low energy.

In this letter, we show that Eq.~(\ref{eq:main}) is in fact consistent with a four-dimensional effective description at low energy, when $\rho_i d\ll 1$, where $d$ is the proper distance between the branes.
This follows because, as we will show, ${\cal C}/a_i^4$ is determined by $\rho_i$, the radion field $d$ and its time-derivative.
At the level of the four-dimensional effective theory, the constant ${\cal C}$ specifies initial conditions for $\rho_i$, $d$ and $\dot{d}$.

Our interpretation has important consequences for nucleosynthesis.
In the era where $\rho_i$ describes radiation, the kinetic energy of the radion quickly redshifts away compared to $\rho_i$ due to Hubble damping.
Thus, within a Hubble time or so, the radion becomes nearly static compared to the expansion rate.
It follows that ${\cal C}/a_i^4$ describes the usual radiation component and the cosmology is conventional.
At matter-radiation equality, however, the radion begins to evolve significantly over a Hubble time.
Therefore, in order to satisfy nucleosynthesis constraints, it suffices that the radion be stabilized before matter-radiation equality.
This is in contrast with previous analyses in which the ${\cal C}/a_i^4$ contribution was neglected~\cite{bine1}.

We begin by reviewing the derivation of the induced Friedmann law on the brane.
The Friedmann equation is then computed at the level of the four-dimensional effective theory and is seen to have the conventional linear dependence on $\rho_i$.
Finally, we show that the two results agree at low energy by explicitly calculating the Weyl tensor contribution in the five-dimensional theory. 

With an empty bulk and ${\bf Z_2}$ symmetry, the action is given by
\begin{equation}
S = \int d^5x \frac{\sqrt{-g}R}{2}  - \sum_{i=1}^2 \int d^4x \sqrt{-g_{(i)}} ({\cal L}_m^{(i)}+2K^{(i)})\,,
\label{eq:action}
\end{equation}
where $g_{(i)}$ and ${\cal L}_m^{(i)}$ are respectively the induced metric and matter Lagrangian density on each brane.
Also, $K_{\alpha\beta}^{(i)} = g^{(i)\;\gamma}_{\;\;\;\alpha}g^{(i)\;\delta}_{\;\;\;\beta}\nabla_{\gamma}n_{\delta}^{(i)}$ is the extrinsic curvature of the $i^{th}$ brane, while $n^{\alpha}_{(i)}$ is the unit normal vector.

The equations of motion derived from~(\ref{eq:action}) consist of the bulk Einstein equations, $G_{\alpha\beta}=0$, supplemented by the Israel matching conditions~\cite{israel,CR}
\begin{equation}
K_{\alpha\beta}^{(i)} = -\frac{1}{2}\left(T_{\alpha\beta}^{(i)}-\frac{1}{3}g_{\alpha\beta}^{(i)}T^{(i)}\right)\,,
\label{eq:israel}
\end{equation}
where $T_{\alpha\beta}^{(i)}$ is the stress-energy tensor associated with ${\cal L}_m^{(i)}$, that is,
\begin{equation}
T_{\alpha\beta}^{(i)} \equiv \frac{2}{\sqrt{-g_{(i)}}}\frac{\delta(\sqrt{-g_{(i)}}{\cal L}_m^{(i)})}{\delta g_{(i)}^{\alpha\beta}}\,.
\end{equation}

Using the Gauss equation, which relates the five-dimensional Riemann tensor to its four-dimensional counterpart induced on the brane, one can derive the four-dimensional Einstein equations on either brane~\cite{shiro,mennim}
\begin{eqnarray}
\nonumber
& & G_{\alpha\beta}^{(i)} = - E_{\alpha\beta}^{(i)} + K^{(i)}K_{\alpha\beta}^{(i)} - g^{\gamma\delta}_{(i)}K_{\alpha\gamma}^{(i)}K_{\beta\delta}^{(i)} \\
& & \;\;\;\;\;\;\;\;\;\;\;\;\;\;\;\;\;\;-\frac{1}{2}g_{\alpha\beta}^{(i)}(K^2_{(i)} - K^{\gamma\delta}_{(i)}K_{\gamma\delta}^{(i)})\,,
\label{eq:ein}
\end{eqnarray}
where $G_{\alpha\beta}^{(i)}$ is the four-dimensional Einstein tensor associated with $g^{(i)}_{\alpha\beta}$, and $E_{\alpha\beta}^{(i)}$ is the electric part of the five-dimensional Weyl tensor $C_{\alpha\beta\gamma\delta}$.
That is,
\begin{equation}
E_{\alpha\beta}^{(i)} \equiv C^{\gamma}_{\;\delta\rho\sigma}n_{\gamma}^{(i)}n^{\rho}_{(i)}g_{\alpha}^{(i)\;\delta}g_{\beta}^{(i)\;\sigma}\,.
\end{equation}
We note in passing that $E_{\alpha\beta}^{(i)}$ is traceless.

Since we are interested in cosmological solutions, the induced metric on either brane is assumed homogeneous, isotropic and spatially-flat.
Thus, we can make the ansatz $ds_{(i)}^2 = g_{\mu\nu}^{(i)}dx^{\mu}dx^{\nu}=-dt^2 + a_i^2(t)d\vec{x}^2$.
Furthermore, we assume that the stress-energy tensor $T_{\mu\nu}^{(i)}$ is that of a perfect fluid with energy density $\rho_i$ and pressure $P_i$.
Eq.~(\ref{eq:ein}) then reduces to
%
\begin{equation}
3H^2_i = - E_{00}^{(i)} + {\cal O}(\rho_i^2)\,,
\label{eq:frw1}
\end{equation}
where $H_i \equiv \dot{a}_i/a_i$ is the Hubble constant on the brane.

The $E_{\alpha\beta}^{(i)}$ tensor is not arbitrary but is constrained by the fact that $G_{\alpha\beta}^{(i)}$ must satisfy the contracted Bianchi identity, $\nabla^{\alpha}G_{\alpha\beta}^{(i)} = 0$.
From Eq.~(\ref{eq:ein}), this implies
\begin{equation}
\nabla^{\alpha}E_{\alpha\beta}^{(i)} = {\cal O}(\rho_i^2)\,.
\end{equation}
To linear order in $\rho_i$, we see that $\nabla^{\alpha}E_{\alpha\beta}^{(i)} \approx 0$.
Thus, $E_{\alpha\beta}^{(i)}$ is approximately conserved and plays the role of a stress-energy tensor in Eq.~(\ref{eq:ein}).
Moreover, since it is traceless, $E_{\alpha\beta}^{(i)}$ is analogous to the stress-energy tensor of a radiation fluid, that is,
\begin{equation}
E_{00}^{(i)} = -\frac{{\cal C}}{a_i^4} + {\cal O}(\rho_i^2)\,,
\label{eq:E001}
\end{equation}
where ${\cal C}$ is a constant.
Substituting in Eq.~(\ref{eq:frw1}), we obtain
\begin{equation}
3H^2_i = \frac{{\cal C}}{a_i^4} + {\cal O}(\rho_i^2)\,.  
\label{eq:rho2}
\end{equation}

Eq.~(\ref{eq:rho2}) displays several apparent peculiarities compared to the usual four-dimensional Friedmann law:
\begin{itemize}
\item As originally noted in Refs.~\cite{lukas2,bine1}, $H^2_i$ depends quadratically on $\rho_i$, rather than linearly.
\item There is a new term, ${\cal C}/a_i^4$, which originates from the bulk Weyl tensor.
Because it decays like $a_i^{-4}$ and does not appear to depend on the brane energy density, this term seems to describe a dark radiation component.
\item The Hubble constant on the first brane, $H_1$, seems independent of the energy density on the other brane, $\rho_2$.
This is contrary to the intuition that gravity should respond to the total stress-energy.
\item The proper distance between the branes does not appear in this expression. 
This is surprising as one would expect that the strength of gravity should vary as the proper distance evolves. 
\end{itemize}

It has been suggested by Cs\'aki {\it et al.}~\cite{csaki} that the unconventional form of the Friedmann law~(\ref{eq:rho2}) is a consequence of requiring the distance between the branes to be static without the addition of a stabilizing potential.
However, in deriving Eq.~(\ref{eq:rho2}) we did not assume that the radion was time-independent.
Moreover, the four-dimensional Friedmann law should hold at low energy, independent of whether the radion evolves with time or not. 
It follows that the solution to the above paradoxes cannot hinge on the existence of a stabilizing mechanism.

We now derive the Friedmann law using the four-dimensional effective theory.
For cosmological applications, the five-dimensional metric is chosen to be isotropic, homogeneous and spatially flat in the three spatial dimensions parallel to the branes.
This corresponds to the ansatz
\begin{equation}
ds^2 = -n^2(t,y)dt^2+a^2(t,y)d\vec{x}^2 + d^2(t,y)dy^2\,,
\label{eq:5dmetric}
\end{equation}
where $0\leq y\leq 1$, with the branes located at $y=0$ and $y=1$.
The Israel conditions~(\ref{eq:israel}) then become
\begin{eqnarray}
\nonumber
& & \left.\frac{a'}{ad}\right\vert_{y=0} = -\frac{\rho_1}{6}\,;\qquad \left.\frac{n'}{nd}\right\vert_{y=0} = \frac{1}{6}(2\rho_1+3P_1)\,, \\
& & \left.\frac{a'}{ad}\right\vert_{y=L} = \frac{\rho_2}{6}\,;\qquad \left.\frac{n'}{nd}\right\vert_{y=L} = -\frac{1}{6}(2\rho_2+3P_2)\,,
\label{eq:israel2}
\end{eqnarray}
where primes denote derivatives with respect to $y$.

In the limit $\rho_id\ll 1$, where the energy is small compared to the mass scale $d^{-1}$ of the Kaluza-Klein modes, the solution to the five-dimensional equations of motion can be written as a perturbative expansion in the small parameter $\rho_id$~\cite{lukas2,lukas}.
We consider the ansatz
\begin{eqnarray}
\nonumber
& & n(t,y) = 1 + \left(\frac{d_0(t)}{6}\right)\sum_{i=1}^2F_i(y)(2\rho_i + 3P_i) \\
& & a(t,y) = a_0(t)\left\{1 -\left(\frac{d_0(t)}{6}\right)\sum_{i=1}^2F_i(y)\rho_i\right\} \label{eq:5d}\\
\nonumber
& & b(t,y) = d_0(t)\left\{1 - \left(\frac{d_0(t)}{6}\right)\sum_{i=1}^2F_i(y)(\rho_i-3P_i)\right\}\,.
\end{eqnarray}
where
\begin{eqnarray}
\nonumber
& & F_1(y) \equiv -\frac{1}{2}y^2 + |y| - \frac{1}{3}\,,\;\;F_2(y) \equiv -\frac{1}{2}y^2+\frac{1}{6}\,.
\end{eqnarray} 
Note that this expression approximately solves the matching conditions~(\ref{eq:israel2}) in the limit $\rho_id_0\ll 1$.

Substituting this ansatz in the five-dimensional equations of motion, $G_{\alpha\beta}=0$, one finds that these reduce to
\begin{equation}
H_0^2 + H_0\left(\frac{\dot{d}_0}{d_0}\right) = \frac{(\rho_1 + \rho_2)}{6d_0} + {\cal O}(\rho_i^2)\,,
\label{eq:4da}
\end{equation}
\begin{equation}
\dot{H}_0 + 2H_0^2 = {\cal O}(\rho_i^2)\,,
\label{eq:4db}
\end{equation}
\begin{equation}
\frac{\ddot{d}_0}{d_0} + 3H_0\left(\frac{\dot{d}_0}{d_0}\right) = \frac{1}{6d_0}\sum_{i=1}^2(\rho_i-3P_i) + {\cal O}(\rho_i^2)\,,
\label{eq:4dc}
\end{equation}
where $H_0(t) \equiv \dot{a}_0/a_0$.
In other words, Eqs.~(\ref{eq:5d}) constitute an approximate five-dimensional solution provided that the functions $a_0(t)$ and $d_0(t)$ satisfy Eqs.~(\ref{eq:4da})-(\ref{eq:4dc}).

It is easily seen that Eqs.~(\ref{eq:4da})-(\ref{eq:4dc}) can alternatively be derived by varying the following four-dimensional effective action
\begin{equation}
S_{eff} = \int_{{\cal M}_4} d^4x \sqrt{-g_4}\left(d_0R_4  - \sum_{i=1}^2{\cal L}_m^{(i)} + {\cal O}(\rho_i^2d_0)\right)\,.
\label{eq:4deff}
\end{equation}
To leading order in $\rho_id_0$, this action is of the Brans-Dicke (BD) form~\cite{bd}, with BD parameter $\omega_{BD} = 0$.
(Note that the kinetic term for $d_0$ can be recovered by going to Einstein frame.)
The four-dimensional metric $g_4$ is given by $ds_4^2 = g^4_{\mu\nu}dx^{\mu}dx^{\nu} = -dt^2 + a_0^2(t)d\vec{x}^2$, and $R_4$ is the corresponding Ricci scalar.

Since the $y$-dependence of the metric~(\ref{eq:5d}) is small for $\rho_id_0\ll 1$, it follows that the induced scale factor on either brane is well-approximated by $a_0(t)$, that is, $a_i\approx a_0$.
In particular, the Friedmann equation on the brane is given by Eq.~(\ref{eq:4da}), and can be written as
\begin{equation}
H_i^2 = \left\{-H_i\left(\frac{\dot{d}_0}{d_0}\right) + \frac{(\rho_1 + \rho_2)}{6d_0}\right\} + {\cal O}(\rho_i^2)\,.
\label{eq:frw2}
\end{equation}
This equation displays all the features one would expect from a four-dimensional Friedmann law.
For instance, $H_i^2$ depends linearly on $\rho_i$, while the gravitational coupling constant weakens as the distance between the branes $d_0$ increases.

While Eqs.~(\ref{eq:frw1}) and~(\ref{eq:frw2}) were obtained using different methods, they should agree at low-energy.
To check this, one can calculate the bulk Weyl tensor corresponding to the metric~(\ref{eq:5d}).
Noting that the unit vector normal to the $y=0$ brane is $n^{\alpha} = (0,0,0,0,d^{-1}(y,t))$, we find, to linear order in $\rho_id_0$,
\begin{equation}
E_{00}^{(i)} \approx \frac{1}{2}\left\{\frac{\ddot{a}_0}{a_0} + \frac{\dot{a}_0}{a_0}\frac{\dot{d}_0}{d_0} - \frac{\dot{a}_0^2}{a_0^2}- \frac{\ddot{d}_0}{d_0}-\sum_{i=1}^2\frac{(\rho_i+P_i)}{2d_0}\right\} \,.
\end{equation}
Using Eqs.~(\ref{eq:4da})-(\ref{eq:4dc}) (which follow from the five-dimensional equations of motion or, equivalently from the four-dimensional effective action) and $H_0\approx H_i$, this reduces to
\begin{equation}
E_{00}^{(i)} = 3H_i\left(\frac{\dot{d}_0}{d_0}\right) - \frac{(\rho_1 + \rho_2)}{2d_0} + {\cal O}(\rho_i^2)\,.
\label{eq:E002}
\end{equation}
Finally, substituting this expression into~(\ref{eq:frw1}) proves that Eqs.~(\ref{eq:frw1}) and~(\ref{eq:frw2}) agree at low energy.

In particular, it is now realized that the apparent puzzles listed below Eq.~(\ref{eq:rho2}) are an artifact of not properly understanding the ${\cal C}/a_i^4$ contribution.
Rather than being a ``dark'' component, we see that this term actually contains all the necessary ingredients to make the Friedmann law on the brane consistent with a four-dimensional effective description.

There is a useful analogy with cosmology in four-dimensional Einstein gravity.
Assuming a constant equation of state $w$, the Friedmann law can be written as
\begin{equation}
3H^2 = 8\pi G_4\left(\frac{{\cal C}'}{a^{3(1+w)}}\right).
\label{eq:f1}
\end{equation}
In analogy with Eq.~(\ref{eq:rho2}), $\rho$ does not appear explicitly in this expression and there is an arbitrary constant ${\cal C}'$.
However, one can use energy conservation, $\dot{\rho}=-3H(1+w)\rho$, to rewrite Eq.~(\ref{eq:f1}) as 
\begin{equation}
3H^2 = 8\pi G_4\rho,
\label{eq:f2}
\end{equation} 
which is the counterpart of Eq.~(\ref{eq:frw2}).
In this form, the dependence on $\rho$ is manifest.

It is remarkable that, to leading order, the cosmological evolution on either brane is that of a radiation-dominated universe for any form of matter on the branes.
This can be understood from the properties of the underlying four-dimensional BD theory~(\ref{eq:4deff}), and is a consequence of the conformal coupling between the radion and the matter fields on the branes.
For any choice of matter content, the radion evolves in such a way that the total stress energy (matter + radion) behaves effectively as radiation.
Explicitly, note that Eq.~(\ref{eq:4db}) implies $H_0=1/(2t)$ to leading order.
It follows that $a_0\sim t^{1/2}$, and therefore $H_0^2\sim 1/a^4_0$.
Comparison with Eq.~(\ref{eq:frw2}) shows that the right-hand side does indeed decay like $1/a^4_0$ for any equation of state.

We have shown that the induced Friedmann law on the brane can be derived from a four-dimensional BD theory at low energy.
If this theory describes the cosmology of our universe, then the measured abundance of light elements requires that Newton's constant $G_4\sim 1/d_0$ vary by less than approximately 10\% from the time of nucleosynthesis until now.
During the radiation-dominated era, the right-hand side of Eq.~(\ref{eq:4dc}) is negligible and the kinetic energy of the radion is damped by the Hubble friction term.
It follows that $d_0$ is nearly static during the radiation-dominated era, and thus the cosmological evolution is conventional.
During the matter-dominated era, however, the right-hand side of Eq.~(\ref{eq:4dc}) cannot be neglected, and it forces the radion to evolve significantly on a Hubble time-scale.
Therefore, in order to satisfy nucleosynthesis constraints, it suffices that the radion be stabilized by the time of matter-radiation equality.

We conclude with a few comments:

\noindent 1. It is often convenient to consider bulk geometries with vanishing Weyl tensor (such as five-dimensional Minkowski or AdS space) since this generally simplifies the five-dimensional equations of motion.
Setting $E_{00}^{(i)}=0$ in Eq.~(\ref{eq:E002}) yields 
\begin{equation}
H_i\left(\frac{\dot{d}_0}{d_0}\right) = \frac{(\rho_1+\rho_2)}{6d_0} + {\cal O}(\rho_i^2)\,.
\label{eq:E00=0a}
\end{equation}
Taking the time-derivative of Eq.~(\ref{eq:E00=0a}) and using Eqs.~(\ref{eq:4da})-(\ref{eq:4dc}), one finds
\begin{equation}
\rho_1 + \rho_2 = {\cal O}(\rho_i^2d_0)\,.
\end{equation}
Thus, we see that requiring the bulk Weyl tensor to vanish implies that the energy density on the branes cancel each other to leading order.

\noindent 2. We have seen that the dynamics are effectively four-dimensional when $\rho_i d\ll 1$.
Suppose that this condition holds initially and that $a_i$ increases with time.
Then, the validity of the four-dimensional effective theory at subsequent times depends on the equation of state on the branes.
Indeed, it can be shown that $\rho_i d\ll 1$ will remain small thereafter provided that $\rho_i\geq -3P_i$ for both branes.
When $\rho_i<-3P_i$, however, the radion increases too fast and $\rho_i d$ eventually becomes of order unity.
One must then solve the full five-dimensional equations of motion to describe the dynamics.

\noindent 3. The above derivation can straightforwardly be generalized to the case of a negative bulk cosmological constant as in RS.
For instance, in the limit $\dot{d}\ll 1$ and $\rho_i \ell\ll 1$, where $d$ is the proper distance between the branes and $\ell$ is the AdS radius, the induced Friedmann law on the positive-tension brane is given by
\begin{equation}
H_1^2 = -\frac{2H_1(\dot{d}/\ell)}{e^{2d/\ell}-1} + \frac{(\dot{d}/\ell)^2}{e^{2d/\ell}-1} + \frac{\rho_1 + \rho_2 e^{-4d/\ell}}{3\ell(1-e^{-2d/\ell})}\,. 
\end{equation}
Note that this reduces to Eq.~(\ref{eq:frw2}) in the limit $d\ll \ell$ where the warp factor is nearly constant.
Details will be presented elsewhere~\cite{future}.

We thank C.P. Burgess, J.M. Cline, S. Gratton, A. Lukas, B.A. Ovrut, U. Seljak, N. Turok, H. Verlinde, D. Waldram, and especially P.J. Steinhardt for helpful discussions.
We also thank P. Bin\'etruy, C. Deffayet and D. Langlois for insightful comments.
This work was supported in part by NSERC (JK), and by an NSF grant NSF-0070928 (RJZ).

\end{multicols}

\begin{thebibliography}{9999}

\bibitem{witten}  P. Ho\v rava and E. Witten, Nucl. Phys. {\bf B460}, 506 (1996); Nucl. Phys. {\bf B475}, 94 (1996).

\bibitem{burt} A. Lukas, B.A. Ovrut, K.S. Stelle and D. Waldram, Phys. Rev. D {\bf 59}, 086001 (1999); Nucl. Phys. {\bf B552}, 246 (1999).

\bibitem{randall} L. Randall and R. Sundrum, Phys. Rev. Lett. {\bf 83}, 3370 (1999); Phys. Rev. Lett. {\bf 83}, 4690 (1999).

\bibitem{shiro} T. Shiromizu, K. Maeda and M. Sasaki, Phys. Rev. D {\bf 62}, 024012 (2000).\label{shiro}

\bibitem{bine2} P. Bin\'etruy, C. Deffayet, U. Ellwanger and D. Langlois, Phys. Lett. B {\bf 477}, 285 (2000).\label{bine2}

\bibitem{shinji} S. Mukohyama, Phys. Lett. B {\bf 473}, 241 (2000).

\bibitem{lukas2} A. Lukas, B.A. Ovrut and D. Waldram, Phys. Rev. D {\bf 61}, 023506 (2000).\label{lukas2}

\bibitem{bine1} P. Bin\'etruy, C. Deffayet and D. Langlois, Nucl. Phys. {\bf B565}, 269 (2000). \label{bine1}

\bibitem{israel}
W. Israel, Nuovo Cim. {\bf B44}, 1 (1966).

\bibitem{CR}
H.A. Chamblin and H.S. Reall, {\it Nucl. Phys.} {\bf B562}, 137 (1999). 

\bibitem{mennim} A. Mennim and R.A. Battye, Class. Quant. Grav. {\bf 18}, 2171 (2001).


\bibitem{csaki} C. Cs\'aki, M. Graesser, L. Randall and J. Terning, Phys. Rev. D {\bf 62}, 045015 (2000).
\label{csaki}


\bibitem{lukas} A. Lukas, B.A. Ovrut and D. Waldram, Nucl. Phys. {\bf B540}, 230 (1999).

\bibitem{bd} C. Brans and C.H. Dicke, Phys. Rev. {\bf 24}, 925 (1961).

\bibitem{future} J. Khoury and R.-J. Zhang, in preparation.


\end{thebibliography}
\end{document}